# The anomalous magnetic properties of CaRuO$_3$ probed by AC and DC magnetic measurements and by low Ti impurity doping

Ivica M. Bradarić [a*], Vladimir M. Matić [a], Ilija Savić [b], Hugo Keller [c]

[a] Institute of Nuclear Sciences "Vinča", University of Belgrade, P.O. Box 522, 11001 Belgrade, Serbia

[b] Faculty of Physics, University of Belgrade, Studentski trg 12, 11001 Belgrade, Serbia

[c] Physik-Institut der Universität Zürich, Winterthurerstrasse 190, CH-8057 Zürich, Switzerland

[*] bradaric@vin.bg.ac.rs                                    December 12, 2017

Calcium ruthenate (CaRuO$_3$) is widely believed to be located close to a quantum critical point due to the strong non-Fermi-liquid behavior expressed in the temperature dependence of electronic transport, specific heat, optical conductivity, etc. However, the corresponding anomalous behavior, marking crossover temperature regimes in the magnetic response of CaRuO$_3$, is still lacking. Here we report detailed AC and DC magnetic susceptibility measurements of CaRuO$_3$ and CaRu$_{0.97}$Ti$_{0.03}$O$_3$. The AC magnetic susceptibility measurements of CaRuO$_3$ show a slight dependence on the frequency of AC magnetic field below $\sim 40$ K, and an additional subtle change of curvature around 12 K. We interpret these results as a critical slowing down of spin fluctuations towards $T = 0$ K. We confirm these observations by magnetic measurements of CaRu$_{0.97}$Ti$_{0.03}$O$_3$, which show a pronounced magnetic response corresponding to the above temperatures.

**I. INTRODUCTION**

Perovskite ruthenates present an attractive playground for the investigation of various ground states of the stoichiometric compounds belonging to the Ruddlesden-Popper (RP) series $A_{n+1}$Ru$_n$O$_{3n+1}$ ($n = 1 - \infty$, $A$ = Ca, Sr) [1]. The interest in these compounds dates from the discovery of superconductivity in the layered quasi-two-dimensional Sr$_2$RuO$_4$ [2], as the Ru-O layers play a crucial role in defining the physical properties of these materials. Therefore, it is possible to tune the effective dimensionality, and consequently the ground state properties, by simply varying the number of layers. Here we focus on the end member of the Ca-RP series, the CaRuO$_3$ compound, and on the influence of the low-Ti doping on its magnetic properties. The pure compound is generally defined as a metallic one and without any detectable magnetic order down to the milli-Kelvin temperature range. However, the metallic behavior is far from being within a classical understanding based on the Landau-Fermi liquid approach. For instance, the DC resistivity increases without saturation in a high-temperature region, thus violating Mott-Ioffe-Regel criterion $k_F l > 1$ (where $k_F$ is the Fermi wavevector, and $l$ is the electron mean free path), whereas at lower temperatures it changes temperature dependence from $\rho(T) \sim T^{1/2}$ to $\sim T^{3/2}$ around 40 K [3]. Magnetic properties are also puzzling. Although CaRuO$_3$ remains a paramagnetic metal down to the lowest temperatures [4], it retains the antiferromagnetic (AFM) fluctuations (negative Curie-Weiss temperature) inherited from the lower dimensional members of the RP series, namely Ca$_2$RuO$_4$ and Ca$_3$Ru$_2$O$_7$, which both settle in the AFM ground state. These observations indicate the proximity of the underlying quantum critical point that governs the anomalous behavior of CaRuO$_3$ [5]. Strangely enough, doping of the isovalent nonmagnetic Ti$^{4+}$ on the Ru$^{4+}$ site in the crystal lattice results in a ferromagnetic-like magnetic response [6-9]. Although the mechanism of the induced ferromagnetism (FM) remains elusive, we elucidate hitherto unreported subtle anomalies in $\chi_{AC}(T)$ (AC magnetic susceptibility dependence on temperature) of pure CaRuO$_3$ and their relation to the corresponding results of the low Ti-doped specimen.

## II. EXPERIMENTAL DETAILS

Polycrystalline samples CaRuO$_3$ and CaRu$_{0.97}$Ti$_{0.03}$O$_3$ were prepared by standard solid-state reaction in the air, starting with the appropriate stoichiometric mixtures of Ru powder + 10% (99.99% Sigma-Aldrich), CaCO$_3$ (99.999% Sigma-Aldrich) and TiO$_2$ (99.995% Sigma-Aldrich) [6-8]. We added a small amount of excess ruthenium powder in order to compensate evaporation during the initial high-temperature reaction. The mixture was preheated at 850$^o$C for 24 h and thus obtained material we mixed again, pressed into pellets and fired at 1000-1200$^o$C for 72 h in air, with two intermediate grindings. Next, we checked the crystal structure of the materials by Philips 1010 powder diffractometer using Cu Kα radiation at room temperature. Magnetic measurements were performed by Physical Properties Measuring System (PPMS), Quantum Design.

## III. RESULTS AND DISCUSSION

Figure 1 shows temperature dependence of DC magnetic susceptibility of CaRuO$_3$ as measured in an applied 2.5 kOe DC magnetic field. Fitting of the Curie-Weiss law $\chi(T) = \chi_0 + C/(T-\theta)$, in the 44 K to 100 K range, yields the effective magnetic moment (estimated from the Curie constant C) $\mu_{eff} = 2.59\ \mu_B$, $\chi_0 = 0.0038\ \text{emu/mol}$, and Curie temperature $\theta = -96\ \text{K}$. These values are in a reasonable agreement with the results that have been reported in the literature [4, 10]. Below $T \sim 40\ \text{K}$ CaRuO$_3$ shows enhanced paramagnetism [11], visible as a deviation from the Curie-Weiss law. To emphasize this anomaly we plotted the difference between $\chi_{DC}$ and $\chi_{CW}$ (up purple triangles).

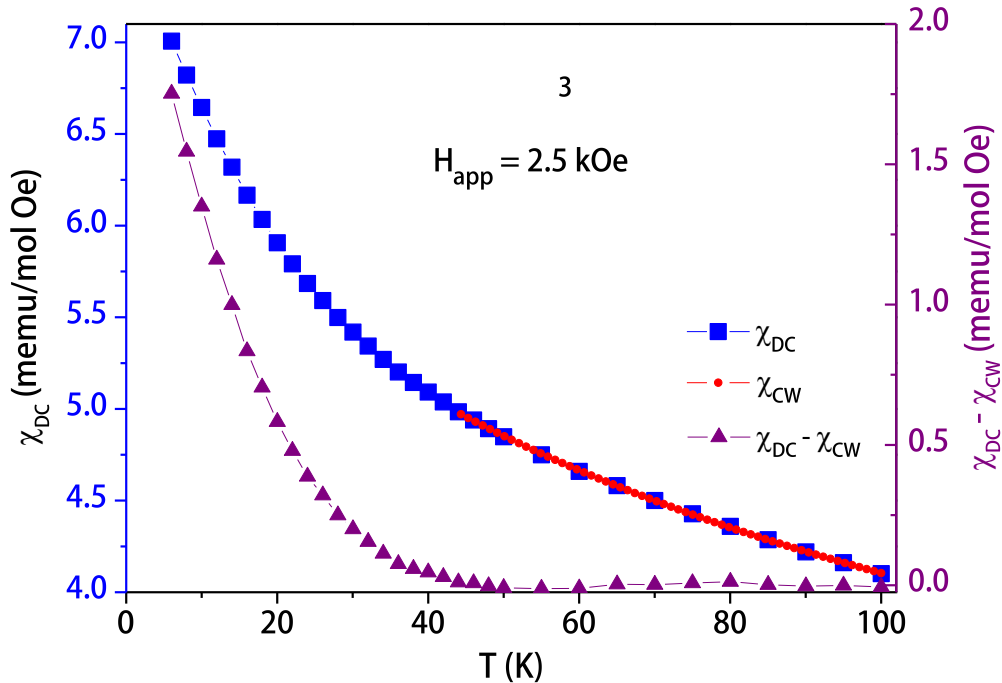

FIG. 1. DC magnetic measurements of CaRuO$_3$. The temperature dependence of DC magnetic susceptibility $\chi_{DC}$ (blue squares), Curie-Weiss law fitting $\chi_{CW}$ (red circles), and the difference $\chi_{DC} - \chi_{CW}$ (up purple triangles).

The increase of the magnetic response is indicative of slowing down of spin fluctuations towards the ordered state. However, as no magnetic order occurs in $CaRuO_3$, it marks the temperature crossover, lending support to the quantum critical scenario [5, 9, 10, 11]. Clearly, slowing down of spin fluctuations reduces the spin-dependent part of electrical resistivity, leading to the steeper decrease with falling temperature. Intriguingly, within this temperature region, anomalous behavior has been reported in the physical properties of $CaRuO_3$ regardless of the form of the compound (ceramic, single crystal or thin single crystalline film), thus pointing to the intrinsic microscopic mechanisms responsible for the observed behaviors [12]. For example, the Hall constant shows a sign reversal from positive to negative with the temperature decrease at $T = 30$ K [4, 9, 13], both magnetic and electronic parts of specific heat develop a broad maximum around 30 K [8, 9], thermoelectric power shows a minimum at 32 K [14], etc.

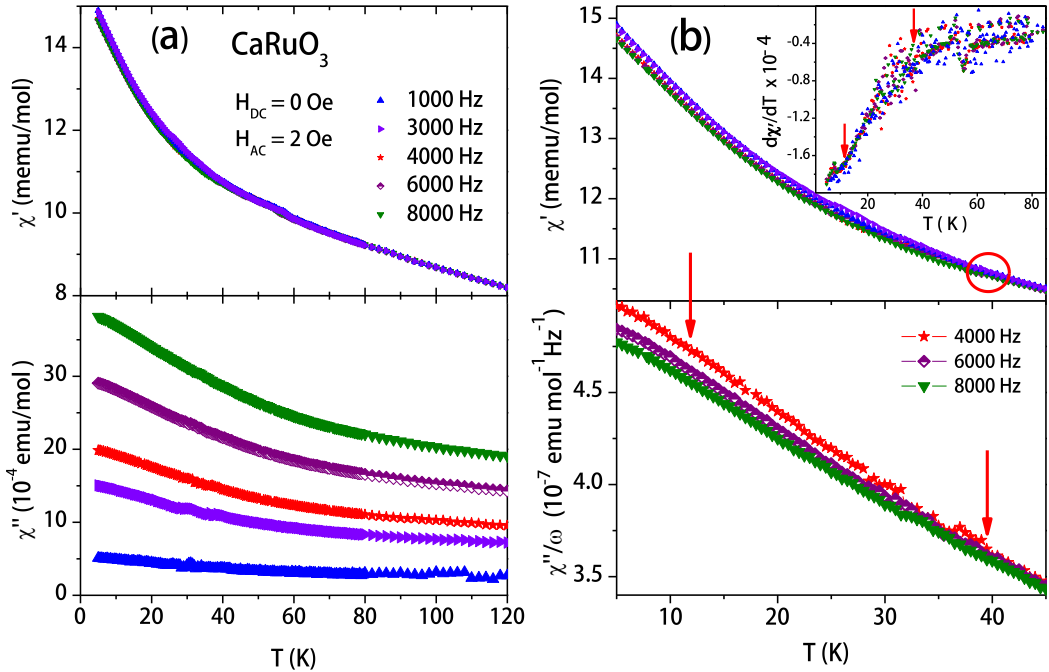

FIG. 2. AC magnetic measurements of $CaRuO_3$. (a) $\chi'(T, \omega)$ the real part of the AC magnetic susceptibility versus temperature (upper panel), and $\chi''(T, \omega)$ the imaginary part of the AC magnetic susceptibility versus temperature (lower panel). (b) An enlarged view of $\chi'(T, \omega)$ (upper panel) and $\chi''/\omega$ (lower panel). The inset in the upper panel shows $d\chi'(T, \omega)/dT$ for various frequencies. The arrows mark anomalies discussed in the main text.

Further, in Fig. 2(a) we display the results of the AC magnetic susceptibility ($\chi_{AC} = \chi' + i\chi''$) dependence on temperature (the upper panel shows the real part and the lower panel shows the imaginary part). Here we observe a noticeable frequency dispersion of $\chi'(T, \omega)$ below 40 K. In order to show this dispersion more clearly, we have redrawn $\chi'(T, \omega)$ within 5-45 K region in the upper panel of Fig. 2(b), and $\chi''/\omega$ in the lower panel. $\chi''$ represents the absorption of the energy due to the induced eddy currents in a metallic material. When the magnetic field penetration depth is larger than the size of the sample, it follows the relation $\chi'' \sim \sigma\cdot\omega$, where $\sigma$ is the electrical conductivity and $\omega$ is the frequency of the AC magnetic field. Apparently, $\chi''/\omega$ curves show the frequency dispersion bellow 40 K. This implies that the additional relaxation processes are present in the system at lower temperatures. Furthermore, there is a slight change of curvature close to 12 K in all $\chi''/\omega$ curves, consistent with changes of slope of $d\chi'/dT$ shown in the inset of the upper panel in Fig. 2(b). We note here again that the observed frequency dispersion in $\chi_{AC}$ reflects the critical slowing down of spin fluctuations above the putative quantum critical point at $T = 0\,\text{K}$.

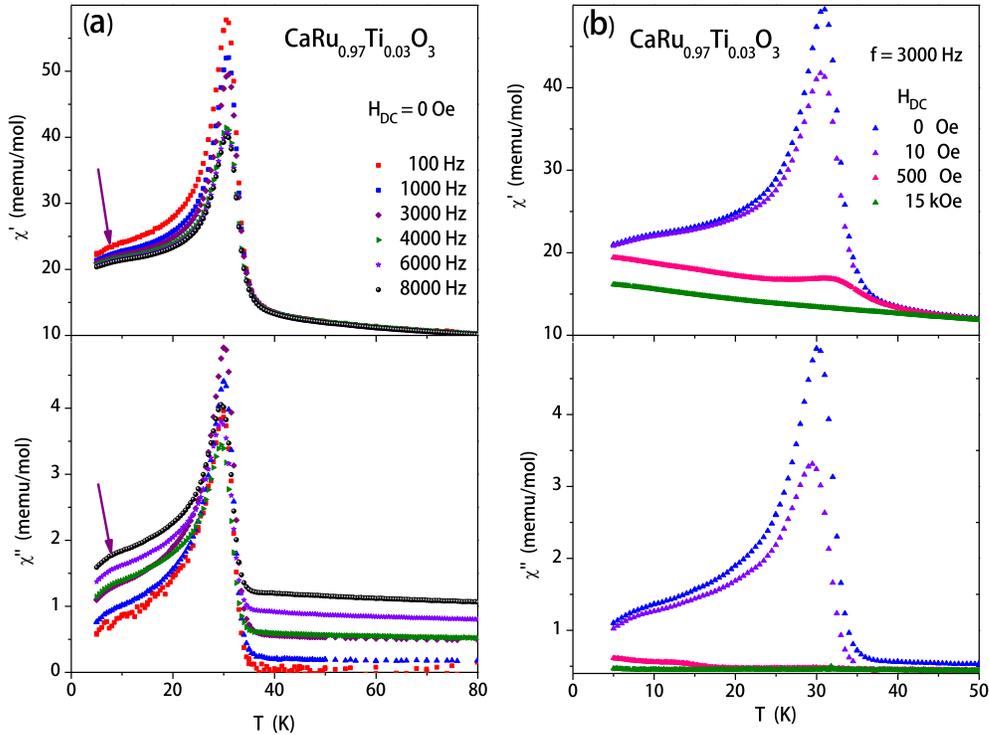

FIG 3. AC magnetic measurements of CaRu$_{0.97}$Ti$_{0.03}$O$_3$. (a) $\chi'(T, \omega)$ (upper panel), and $\chi''(T, \omega)$ (lower panel). Arrows mark the low-temperature anomaly. (b) $\chi'(T, \omega = 3\,\text{kHz})$ (upper panel), and $\chi''(T, \omega = 3\,\text{kHz})$ (lower panel) measured in various applied DC magnetic fields.

Now we turn to the analysis of magnetic properties of the low Ti-doped CaRuO$_3$. In Fig. 3(a) we present the results of AC magnetic measurements of the CaRu$_{0.97}$Ti$_{0.03}$O$_3$ compound. Evidently, a strong peak in $\chi_{AC}(T, \omega)$ develops around 30 K, signaling the ferromagnetic order in this system. In addition, another anomaly is clearly visible around 10 K in the real and imaginary parts of $\chi_{AC}(T, \omega)$. In the following, we shall show that these phenomena are closely related to the behaviors of the pure compound. The frequency dependence of position and the intensity of the main peak and the low-temperature anomaly are clearly visible. Next, we observe the strong dependence of the intensity of the main peak on the applied DC magnetic field, as shown in Fig. 3(b). These dependencies are commonly associated with the spin glass behavior [15]. Indeed, in Fig. 4(a) we show the detailed frequency dependence of the temperature position of the main peak centered around 30 K. In Fig. 4(b) and Fig. 4(c) we present the results of the fitting to the dynamical scaling law $\omega = \omega_0 \left[ (T_f - T_g)/T_g \right]^{z\nu}$ for the main peak and the low-temperature anomaly, respectively. Here, $\omega_0$ is the microscopic relaxation frequency, $T_f$ is the frequency-dependent freezing temperature defined as the maximum in $\chi_{AC}$, $z$ is the dynamic exponent, and $\nu$ is the correlation length exponent. The obtained fitting parameters for the main peak $\omega_0 = 8.7 \times 10^{11}$ Hz, $z\nu = 4.2$, and $T_g = 30.3$ K, are typical values observed in the canonical spin glasses. Moreover, the frequency shift in $T_f$, $\Delta T_f / \left[ T_f \Delta(\log \omega) \right] = 0.002$, is also consistent with the corresponding values of the canonical spin glasses. We have determined the frequency dependence of the low-temperature anomaly by finding the intersection point between the temperature axis and $d^2\chi_{AC}(T,\omega)/dT^2$. The fitting parameters for the low temperature anomaly are $\omega_0 = 5 \times 10^6$ Hz, $z\nu = 4.7$, $T_g = 9.7$ K, and $\Delta T_f / \left[ T_f \Delta(\log \omega) \right] = 0.07$. These results are also in accordance with the spin glass behavior, although the microscopic relaxation frequency shows much slower dynamics.

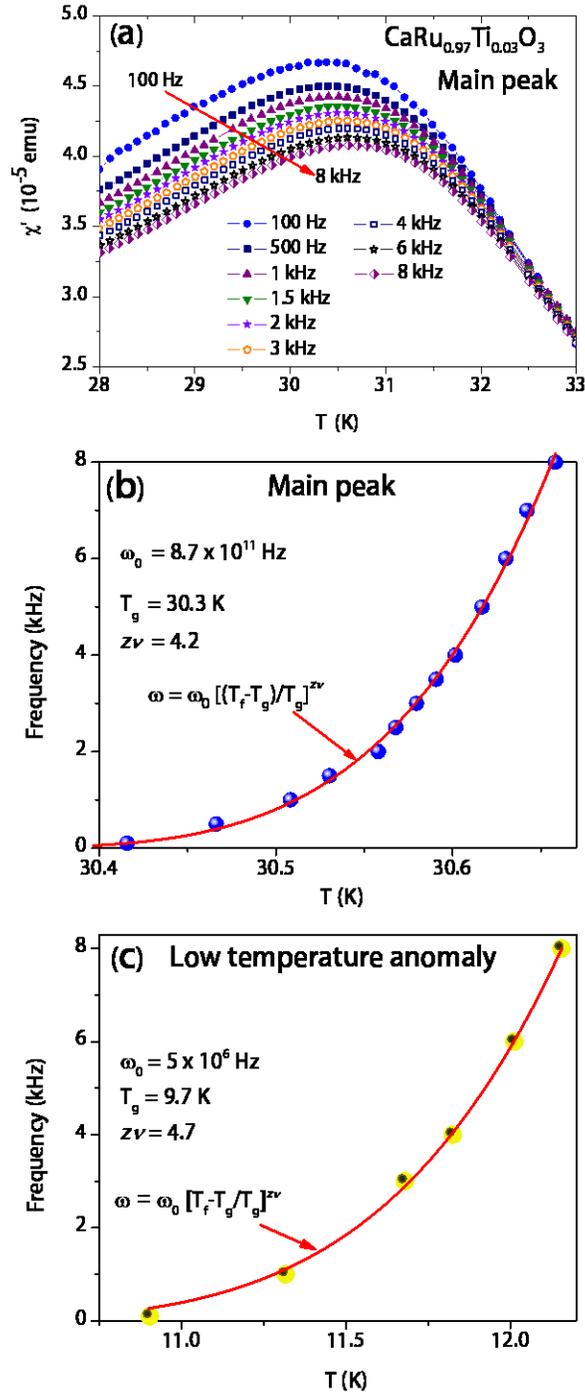

FIG. 4. Analysis of the frequency dependence of the main peak around 30 K and the low-temperature anomaly around 10 K. (a) Detailed $\chi'(T, \omega)$ measurements of the main peak of CaRu$_{0.97}$Ti$_{0.03}$O$_3$, from $T$ = 28 K to 33 K and $\omega$ = 100 Hz to 8 kHz. The results of fitting of the dynamic scaling law (red line) to the experimental data (blue circles) for (b) the main peak and (c) the low-temperature anomaly.

Figure 5(a) shows the results of the temperature dependence of magnetization at various DC magnetic fields. The divergence point between zero-field-cooled (ZFC) and field-cooled (FC) branches rapidly moves to lower temperatures with an increasing magnetic field. Such behavior is often attributed to the spin glass characteristics, but we note that the FC branch in spin glasses shows temperature independent behavior below the divergence point, which is not the case here. Magnetization curves in Fig. 5(b) show paramagnetic behavior for $T \geq 50$ K, and an opening of the hysteresis loop at 5 K, with coercive field $H_c = 1.19$ kOe, remanent magnetization $M_r = 0.017$ emu, and the saturation above 10 kOe. These observations point to the ferromagnetic ordering within an inhomogeneous system, which leads us to the construction of the following physical picture. Namely, introducing a nonmagnetic $Ti^{4+}$ ($3d^0$, $S=0$) impurity into the paramagnetic metal $CaRuO_3$, where $Ru^{4+}$ ($4d^4$: $t_{2g}^4$, $e_g^0$ with total spin $S=1$), triggers a ferromagnetic-like ordering around 35 K [6-9]. This means that charge carriers in the neighborhood of the Ti impurity acquire spin polarization, thus forming a "droplet" of magnetic order. Further, this allows us to view the Ti impurities as effectively ferromagnetic impurities embedded in the metallic paramagnetic background. Now, it immediately follows that ferromagnetic droplets mutually interact via long-ranged indirect-exchange RKKY (Ruderman, Kittel, Kasuya, Yosida) interaction, whose Hamiltonian is $H = J(r)\mathbf{S}_i\mathbf{S}_j$. Here $J(r)$ is the RKKY interaction, $\mathbf{S}_i$ and $\mathbf{S}_j$ are the magnetic moments of the droplets, and $r$ is the distance between them. Because the RKKY interaction proceeds via coupling of local magnetic moments with spins of itinerant electrons, it causes damped oscillations in the susceptibility of charge carriers and therefore mixed ferromagnetic and antiferromagnetic couplings between magnetic impurities, depending on their separation. This establishes the ground for the observed spin-glass behavior in the AC magnetic susceptibility measurements. Next, as spin fluctuations in pure $CaRuO_3$ slow down below 40 K, it is easier to polarize more spins around Ti impurities, leading to the growth of ferromagnetically ordered droplets. Finally, $CaRuO_3$ shows paramagnetic uniaxial anisotropy [16], so that the order parameter belongs to the three-dimensional Ising type symmetry. The physical picture outlined above closely corresponds to the theoretical models developed to describe the physics in systems with the quenched disorder [17, 18]. In short, the theoretical model for a three-dimensional infinite-range Ising model with planar defects (disordered layers or grain boundaries) predicts smearing of the sharp continuous phase transition due to development of the static long-range order on "rare regions" (regions that are devoid of impurities). Accordingly, the order parameter is inhomogeneous in space near a smeared phase transition. In Fig. 5(c) we plot $M(T)$ ($H_{DC} = 100$ Oe) and $\chi_{AC}(T, \omega = 1\,\text{kHz})$, and in Fig. 5(d) we present the fitting results of the theoretical expression $M \sim exp\left(-B|T_c - T|^{-1/2}\right)$, $(T \to T_c-$, $T_c$ is the critical temperature of the clean system), to the $M(T)$ ($H_{DC} = 100$ Oe) experimental data.

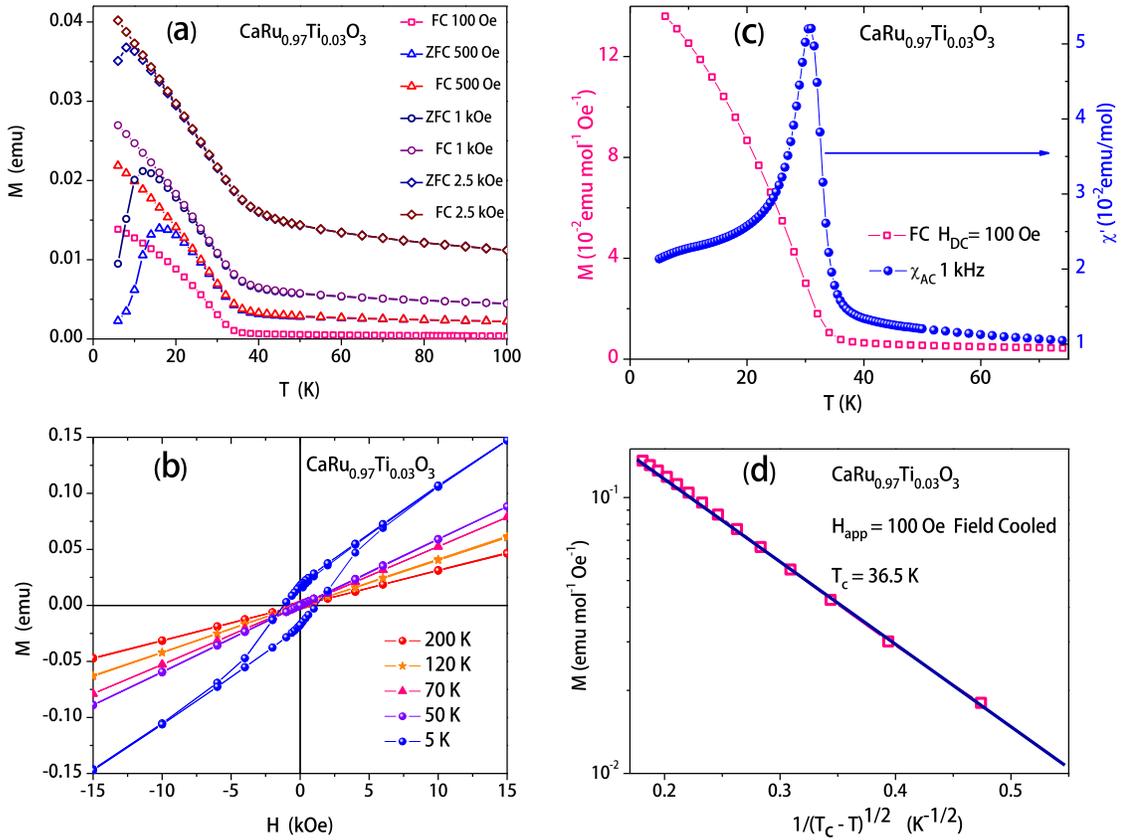

FIG. 5. Magnetization measurements of CaRu$_{0.97}$Ti$_{0.03}$O$_3$ and analysis. (a) Zero-field-cooled and field-cooled DC magnetization dependence on temperature at various applied magnetic fields. (b) Magnetic hysteresis loops at various temperatures. (c) Field–cooled DC magnetic susceptibility versus temperature measured in $H_{DC} = 100$ Oe magnetic field (left axis) and $\chi'(T, \omega = 1\,\text{kHz})$ (right axis). (d) The field-cooled DC magnetic susceptibility as a function of $(T_c - T)^{-1/2}$, open red squares are experimental data and the straight line is the fitting result.

We note here, however, that in our case, "rare regions" are located exclusively around Ti impurities, and that the critical temperature does not depend on their concentration. Apparently, the excellent fitting results justify the physical picture proposed above. This brings us to the interpretation of the observed coinciding anomalies in the pure and doped compounds. Namely, because the RKKY interaction depends on spin dynamics of itinerant electrons, any change in their state will necessarily affect the magnetic susceptibility behavior in $CaRu_{0.97}Ti_{0.03}O_3$. Therefore, we argue that both temperature anomalies in Ti-doped samples are directly related to the crossover temperatures in the pure compound. Indeed, around 10 K (low-temperature glassy behavior in the Ti-doped sample) a number of physical quantities of pure $CaRuO_3$ show pronounced changes. Below 10 K specific heat shows logarithmic temperature divergence [5, 9, 19], neutron scattering measurements show significant slowing down of spin fluctuations [10], magnetoresistance shows change from negative to positive [3], and the Hall constant shows a minimum and a slight increase towards lower temperatures [4]. Interestingly, the applied static magnetic field reduces the specific heat divergence, which taken together with positive magnetoresistance below 10 K, point to the recovery of the Fermi surface. Indeed, the recent electric transport measurements performed on the high-quality thin film $CaRuO_3$ present the evidence of a fragile Fermi liquid ground state below $T_{FL}$ = 1.5 K [20]. In addition, $T_{FL}$ strongly increases with magnetic field thereby suggesting a suppression of the spin fluctuations, in accordance with other experiments. Hence, the effect of Ti impurity is similar to the effect of the applied static magnetic field, indicating the proximity of $CaRuO_3$ to a zero-magnetic-field quantum critical point.

## IV. CONCLUSION

In this article we have provided evidence for several novel features in the metallic quantum paramagnet $CaRuO_3$ and in the low Ti-doped compound $CaRu_{0.97}Ti_{0.03}O_3$. AC magnetic susceptibility measurements of $CaRuO_3$ showed slight frequency dependence below 40 K, visible in the real and the imaginary parts, that coincides with the crossover temperature dependence of other physical properties reported in the literature. In addition, we have reported a low-temperature anomaly visible as a slight change of curvature in $\chi_{AC}(T,\omega)$. These results indicate a progressive slowing down of spin fluctuations towards the lower temperatures. The results of magnetic measurements of $CaRu_{0.97}Ti_{0.03}O_3$ show the close correspondence with the anomalies recorded in pure $CaRuO_3$, that is, we have observed a spin glass behavior at 30.3 K and 9.7 K. We have explained these phenomena by assuming the RKKY interaction between ferromagnetic islands (around Ti impurities) immersed in a metallic paramagnetic background. Finally, we have confirmed this physical picture by finding an excellent agreement between our experimental data and the theoretical model of the smeared phase transition in Ising systems with planar defects.

## ACKNOWLEDGMENTS

This work was financially supported by the Ministry of education, science, and technological development, Republic of Serbia (Grant No. 171027) and the Swiss National Science Foundation.